\documentclass[prl,twocolumn,showpacs,preprintnumbers,amsmath,amssymb]{revtex4}
\usepackage{epsf}
\usepackage{amsfonts}
\usepackage{amssymb}
\usepackage{amsthm}
\usepackage{graphicx}
\newcommand{\cf}[1]{\langle #1 \rangle}                      
\newcommand{\bra}[1]{\langle #1 \!\mid\!}                    
\newcommand{\ket}[1]{\!\mid\! #1 \rangle}                    
\newcommand{\phdagger}{\mathop{\phantom{\dagger}}}           
\newcommand{\cop}[1]{c^{\phdagger}_{#1}}                     
\newcommand{\cdop}[1]{c^{\dagger}_{#1}}                      

\newcommand{\bml}{\begin{mathletters}}
\newcommand{\eml}{\end{mathletters} \hspace{-5pt}}

\begin{document}

\title{Entanglement Scaling in the One-Dimensional Hubbard Model at Criticality}

\author{Daniel Larsson and Henrik Johannesson}

\affiliation{Department of Physics, G\"oteborg University, SE 412 96 G\"oteborg, Sweden}


\begin{abstract}
We derive exact expressions for the local entanglement entropy ${\cal E}$
in the ground state of the one-dimensional Hubbard model at a quantum phase transition
driven by a change in magnetic field $h$ or chemical potential $\mu$. The
leading divergences of $\partial{\cal E}/\partial h$ and $\partial{\cal E}/\partial \mu$ 
are shown to be
directly related to those of the zero-temperature spin and charge susceptibilities.  
Logarithmic corrections to scaling signal a change in the number of local states
accessible to the system as it undergoes the transition.   
\end{abstract}


\maketitle

\newpage


Entanglement is a generic feature of quantum systems, implying the possible existence of non-local
correlations. Such correlations $-$ which lead to highly counter-intuitive phenomena $-$ were long seen as
an artifact of quantum mechanics \cite{EPR}. With the advent of quantum information theory it is now
understood
that entanglement $-$ and the correlations associated with it $-$ is not only intrinsic to the fabric of
reality \cite{Bell}, but can also be used as a physical resource, essential for performing such tasks as
teleportation or quantum computing \cite{NielsenChuang}. 

A new line of research \cite{Osterloh, Osborne}
points to a connection between the 
entanglement of a many-particle
system $-$ as quantified by a properly chosen measure $-$ and the appearance of a (zero-temperature)
quantum
phase transition (QPT) \cite{Sachdev}. Barring accidental occurrences of
non-analyticity, a discontinuity [singularity] in
the [derivative of the] ground state concurrence of an $N$-qubit system appears to be
associated 
with a first [second] order QPT 
\cite{Wu} (with {\em concurrence} measuring the entanglement between two neighboring
qubits
\cite{Wootters}). These and related results are important as they hold promise of
novel perspectives on condensed matter, drawing on insights from quantum information theory. 
By analyzing entanglement properties one expects to gain insight into how the associated
non-local (purely quantum) correlations influence the critical behavior of a quantum phase transition. Building an
understanding of this connection should enable breakthroughs in the design of future experimental probes of collective
quantum phenomena.
Also, 
architectures for quantum information processing that take advantage of the entanglement in the vicinity of a
quantum phase transition {\em (quantum adiabatic computing)} \cite{Orus} should benefit from a detailed understanding of
entanglement scaling properties. 

Most results to date on the {\em entanglement - QPT connection} have been obtained from numerical
studies of finite lattice spin systems, supplemented by some analytical results \cite{Latorre}. 
Much less is known about entanglement scaling properties of {\em itinerant} electron systems. 
In this Letter
we make a dent on this important problem by studying the
one-dimensional Hubbard model close to a quantum phase transition. Recent work on this and related 
models show that features of the ground state phase diagram can be reproduced
by studying certain characteristics of the local entanglement entropy \cite{Korepin,Gu,Anfossi1,Anfossi2,Wang}.
Here we exploit the {\em
Bethe
Ansatz} solvability of the Hubbard model to derive exact expressions for the critical scaling of the 
local entanglement entropy 
${\cal E}(\psi_0)$ of its ground state $\ket{\!\psi_0}$ as function
of magnetic field $h$ and chemical potential $\mu$. We find that the leading scaling behavior of $\partial
{\cal E}/\partial \mu $
for repulsive interaction coincides with that of the charge susceptibility $\chi_C$.
A similar result holds for $\partial {\cal E}/\partial h$,
but with logarithmic corrections that signal a change of
dimension of the accessible local state space at the transition. The fact that an entanglement 
measure of a critical many-particle
system can be {\em quantitatively} linked to a physical observable is a striking result, and goes beyond
standard constructions of entanglement witnesses \cite{Terhal} that merely detect the presence of
entanglement. To what extent our results can be generalized to other quantum systems is yet to be answered.

To set the stage, let us recall that the very notion of entanglement of a composite quantum system relies
on the
tensor product structure of its Hilbert space. When the system is made up of itinerant electrons, however,
the physical
subspace is restricted to an anti-symmetrized one which lacks a natural product structure.  
One may circumvent the problem by passing to an occupation number  
representation spanned by the $4^L$ basis states $\ket{\!n}_1 \otimes \ket{\!n}_2 \otimes
 ... \otimes \ket{\!n}_L $,
where, in obvious notation, $\ket{\!n}_j = \,\ket{0}_j, \, \ket{\,\uparrow}_j, \,\ket{\,\downarrow}_j$, or
$\ket{\,\uparrow \downarrow}_j$
is a local state at site $j$, with $L$ the number of sites on the lattice \cite{Zanardi}. This is a
convenient basis
in which the tensor product structure is manifestly recovered, with the local states describing {\em
electronic modes}
easily accessible to an observer. 
By splitting the system into two
parts $A$ and $B$ one can then proceed as usual and define the {\em entanglement entropy} ${\cal E}$ of a
pure state $\ket{\psi}$ as ${\cal E} = - \mbox{Tr}(\rho_A \mbox{log}_2 \rho_A)$ \cite{NielsenChuang}.
The reduced density matrix $\rho_A = \mbox{Tr}_B(\rho)$ is obtained from the full density matrix $\rho
=\ \ket{\psi} \bra{\psi}$ by tracing out 
the local states belonging to $B$.
In what follows we focus on the entanglement entropy of a single site, obtained by taking
$A$ to be a single (arbitrarily chosen) site, with $B$ the rest of the system.   

We begin by studying the Hubbard model with an applied magnetic field $H$:
\begin{equation} \label{HubbardHamiltonian}
{\cal H} = -t \sum^L_{j=1 \atop \delta = \pm 1} 
\cdop{j \alpha} \cop{j+\delta \alpha} + U \sum^L_{j=1} n_{j \uparrow} n_{j \downarrow}
- \mu_B H \sum^L_{j=1} S^z_j.
\end{equation}
Here $\cdop{j \alpha}$ and $\cop{j \alpha}$ are creation and annihilation operators for electrons at site $j$ with
spin $\alpha = \uparrow, \downarrow$, and
$n_{j \alpha} \equiv \cdop{j \alpha} \cop{j \alpha}$ and $S^z_j = (n_{j \uparrow} - n_{j \downarrow})/2$ are 
the corresponding number and spin operators, respectively. We assume periodic boundary conditions.
In the following we shall work with dimensionless quantities $u \equiv U/4t$ and $h \equiv \mu_B H/t$,
putting $t=1$ (of dimension energy). 
Using the fact that the Hamiltonian in (\ref{HubbardHamiltonian}) is
translational invariant and conserves
particle number as well as the z-component of the total spin, it is easy to verify that the reduced
density matrix
$\rho_A$ for a single site $A$ is diagonal in the chosen basis. It follows 
that the 
corresponding single-site entanglement of the ground state $\ket{\psi_0}$ can be written as
\begin{equation} \label{OneTangle}
{\cal E} \!=\! -w_0 \log_2 w_0 - w_{\uparrow} \log_2 w_{\uparrow}-
w_{\downarrow}\log_2 w_{\downarrow} - w_2 \log_2 w_2 
\end{equation}
with
\begin{eqnarray} \label{Parameters}
w_2 & = & \cf{n_{j \uparrow} n_{j \downarrow}}_0, \ \ \ \
w_{\alpha}  =  \cf{n_{j \alpha}}_0 - w_2, \   \alpha = \uparrow, \downarrow  \nonumber \\
w_0 & = & 1 - w_{\uparrow} - w_{\downarrow} - w_2. 
\end{eqnarray}
The problem is thus reduced to calculating the expectation values for double and single (spin-up and
spin-down)
occupancies in the ground state.

\begin{figure}[tbh]
\begin{center}
\includegraphics[width=0.35\textwidth]{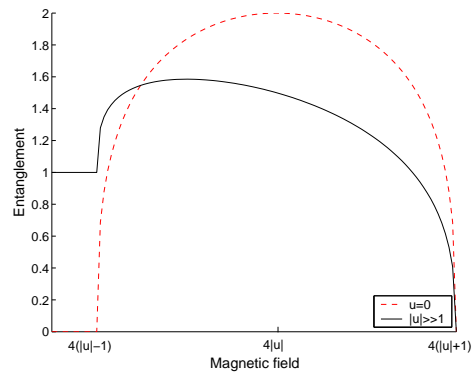}
\caption[Site-tangle versus magnetic field.]{Entanglement entropy ${\cal E}$ of a single site versus magnetic field $h$
for the attractive Hubbard model with $|u|\gg 1$ (solid curve). 
For comparison, the single-site entanglement for the free case ($u=0$) is shown by the dotted curve
(on a different scale).} 
\label{MagField}
\end{center}
\end{figure}

Let us first look at the case of attractive interaction, $u\!<\!0$, with $n\!=\!1$ {\em (half-filling)}.
In the limit $|u|\!\gg\!1$
we can use the Hellman-Feynman theorem, $\cf{\partial H / \partial u}_0 = \partial E_0/ \partial u$,
together with
the known {\em Bethe Ansatz} result for the ground state energy \cite{Yang}, $E_0/4L = u(1/2 - m)
-(1/2\pi)\sin (2\pi m) 
+ {\cal O}(1/u)$ to obtain
\begin{equation}  \label{w}
w_2 = \frac{1}{4L} \frac{\partial E_0}{\partial u} = \frac{1}{2} - m + {\cal O}(1/u^2),
\end{equation}
with $m = (1/2L)\sum_{j=1}^L \cf{n_{j \uparrow} - n_{j \downarrow}}$ the magnetization per site. 
Neglecting the ${\cal O}(1/u^2)$ corrections it follows immediately from (\ref{Parameters}) and (\ref{w}) that
\begin{equation} \label{uANDz}
w_{\uparrow} = 2m, \ \ \ w_{\downarrow} = 0, \ \ \ w_0 = \frac{1}{2} - m, \ \ h \ge 0.   
\end{equation}
Combining (\ref{OneTangle}), (\ref{w}), and (\ref{uANDz}), we obtain for the single-site entanglement:
\begin{equation} \label{AttractiveOneTangle}
{\cal E} = 
-2m\log_2(2m) - (1-2m)\log_2 (\frac{1}{2}-m), \ \ h \ge 0.
\end{equation}
The dependence of the magnetization on the applied field can also be derived from the
ground state energy, and one finds
\begin{eqnarray}  \label{Magnetization}
m(h)= \left\{\begin{array}{ll}
0&0\leq h<h_{c1}\\
\frac{1}{2\pi}\arccos\left(-(u+\frac{h}{4})\right)&h_{c1}\leq h\leq h_{c2}\\
\frac{1}{2}&h_{c2}<h\end{array}\right.
\end{eqnarray}
with lower [upper] critical field $h_{c1} =  4(|u| - 1) \ [h_{c2} = 4(|u| + 1)]$.
The single-site entanglement as a function of magnetic field, ${\cal E} = {\cal E}(h)$, can now be read off
from (\ref{AttractiveOneTangle}) and (\ref{Magnetization}). The result 
for the $|u|\gg 1$ limit is plotted in Fig. (\ref{MagField})
for large values of $h$. 
Note that in this limit there are two local states,
$\ket{0}$ and $\ket{\,\uparrow \downarrow}$, available to the system
when $h < h_{c1}$, implying that ${\cal E}(h) = 1$. 
In contrast, the fully magnetized state for $h > h_{c2}$ is a direct product of local spin-up states, 
and hence ${\cal E}(h) = 0$.
For comparison we have plotted the single-site entanglement for free electrons also in Fig. (\ref{MagField})
(for both positive and negative values of the magnetic field). 
This result is easily obtained from 
Ref. \cite{Yang} 
by noting that $w_2 = 1/4 - m^2$ when $u\!=\!0$, with $m=(1/\pi)\arcsin(h/4)$ in the interval 
$-4\!<\!h\!<\!4$.

The phase transitions at $h_{c1}$ and $h_{c2}$ are second order, with diverging spin susceptibilities 
$\chi_{Si} = (32\pi^2|h-h_{ci}|)^{-1/2}, i=1,2$ \cite{Yang}. 
As mentioned in the introduction, in the case of an $N$-qubit system there is
strong evidence that the
derivative of the ground state concurrence with respect to a critical parameter 
diverges at a second-order QPT 
\cite{Wu}. 
For the present problem (where the local degrees of freedom have {\em four} components,
not two as for a qubit)
the plot in Fig. \!(\ref{MagField}) suggests a divergence of 
$\partial {\cal E} / \partial h$ as $h \rightarrow h_{c1 +}$ and $h \rightarrow h_{c2 -}$.   
To analytically check whether the derivative of the single-site entanglement 
is a true marker of quantum criticality for this problem, 
we write $u+h/4 = (h-h_{ci})/4 +(-1)^i, i=1,2$ and expand $\partial{\cal E}/ \partial h$ in $h-h_{c1}$
and
$h_{c2}-h$, respectively. We obtain
\begin{equation} \label{OneTangleGradient1}
\frac{\partial{\cal E}}{\partial h} = (-1)^i\frac{\chi_{Si}}{\ln(2)}(\ln|h-h_{ci}| +
\mbox{const.}), \ \ i=1,2 \ \ \
\end{equation}
for $h \rightarrow h_{c1 +}$ and $h \rightarrow h_{c2 -}$, respectively.  This confirms that
$\partial {\cal E}/\partial h$ diverges at the magnetic phase transitions. Moreover, it shows that  
{\em the divergence of} $\partial {\cal E}/\partial h$ {\em is given by the spin susceptibility} $-$ up to a
logarithmic correction. By writing the leading 
term of $\partial {\cal E} / \partial h$ on the ''mixed'' form $ (-\partial m / \partial h)
(2\ln(w_{\uparrow}) - \ln(w_0)
- \ln(w_2))/\ln 2$ [cf. Eqs. (\ref{OneTangle}) and (\ref{AttractiveOneTangle})], and combining 
this expression with (\ref{Magnetization}), one sees
that the logarithmic divergence in (\ref{OneTangleGradient1}) comes from a change of the 
number of local states accessible to the system as it undergoes the transition:  As $h \rightarrow h_{c1
+}$ the local
spin-up states get suppressed $(w_{\uparrow}\!\rightarrow\!0$), while for $h \rightarrow h_{c2 -}$ 
both empty and doubly occupied local states 
get suppressed ($w_0, w_2 \rightarrow 0$). 

Turning to the half-filled case with repulsive interaction, $u>0$, a QPT now occurs
only at the value of 
the field for which the magnetization saturates: $h_{c2} = 4(\sqrt{1+u^2} -u)$ \cite{TakahashiBook}.
As shown by Takahashi, the ground state
energy for {\em any} finite value of $u>0$ in the critical region $h \rightarrow h_{c2-}$ can be expanded in
terms of the  
expectation value for single spin-down occupancy \cite{Takahashi}:
\begin{eqnarray} \label{RepulsiveEnergy}
\frac{E_0}{4L} & = & -\large(\sqrt{1 + u^2} - u\large)\cf{n_{j \downarrow}}_0  \nonumber \\
              & + & \frac{\pi^2}{24} \frac{1}{\sqrt{1 + u^2}}\cf{n_{j \downarrow}}_0^3 + 
{\cal O}(\cf{n_{j\downarrow}}_0^4).
\end{eqnarray}
With the same procedure as used for the attractive case above, 
Eq. (\ref{RepulsiveEnergy}), together with (\ref{OneTangle}) and (\ref{Parameters}),
yield: 
\begin{equation}  \label{RepulsiveEntanglement}
\frac{\partial {\cal E}}{\partial h} = \frac{C}{2\ln(2)} \chi_S \,(\ln |h - h_{c2}| +
\mbox{const.}), \ \
\
 h \rightarrow h_{c2 -}.
\end{equation}
Here $C = 2 - u/\sqrt{1 + u^2},$ and $2\pi \chi_S =
(4+4u^2)^{1/4}|h-h_{c2}|^{-1/2}$. 
The logarithmic correction in (\ref{RepulsiveEntanglement}) now signals the suppression of all but the
spin-up
states as one approaches the saturation point $h_{c2}$ from below. 

\begin{figure}[tbh]
\begin{center}
\includegraphics[width=0.35\textwidth]{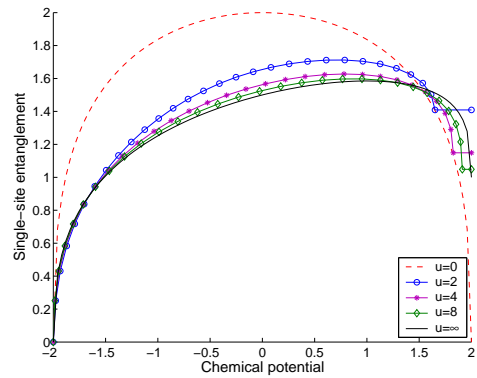}
\caption[Site-tangle versus chemical potential.]{Entanglement entropy ${\cal E}$ of a single site versus chemical potential 
$\mu$ for the repulsive
Hubbard model.
The plateaus correspond to half-filling $(n=1)$, cut off at $\mu=2$. The dotted curve is that for free electrons $(u=0)$, plotted in the region $0\le n\le 2$.}
\label{ChemPot}
\end{center}
\end{figure}

We next study the effect of a varying chemical potential on the single-site entanglement of an {\em open}
system. We make the simplifying assumption that the environment acts solely as a particle reservoir \cite{Note}, 
and add the term ${\cal H}_{\mu} = - \mu \sum_{j=1}^L (n_{j \uparrow} + n_{j \downarrow})$ 
to the Hamiltonian in (\ref{HubbardHamiltonian}), with $\mu$ a dimensionless
chemical potential (multiplied by the hopping amplitude $t=1$).
To simplify further we turn off the magnetic field in (\ref{HubbardHamiltonian}), putting $h=0$.
Focusing on the case of repulsive interaction, $u>0$, with $n\le 1$, the
system exhibits two quantum critical points \cite{LiebWu}: $\mu_{c1} = -2$ and
$\mu_{c2}\!=\!2-4\int_0^{\infty} \!J_1(\omega)(\omega [1 + \exp(2\omega u)])^{-1}d\omega$, with 
$J_1(\omega)$ a first-order Bessel function.
Both transitions are second-order with diverging charge susceptibilities $\chi_{Ci} =
c(u)|\mu-\mu_{ci}|^{-1/2}, \
i=1,2$ in the limits $\mu \rightarrow \mu_{c1 +}$ {\em (empty lattice transition)} and 
$\mu \rightarrow \mu_{c2 -}$ {\em (Mott transition)}, respectively (with $c(u)$ a
positive $u$-dependent constant). 
To obtain the single-site entanglement ${\cal E}$ we first notice that
${\cal H}_{\mu}$ conserves spin and particle number for fixed $\mu$, and that hence the expression for 
${\cal E}$ in (\ref{OneTangle}) remains valid. Recalling from the Lieb-Mattis theorem \cite{LiebMattis}
that the ground state has zero spin (for any $n$ with
$nL$ an even
integer) we can write the parameters appearing in (\ref{OneTangle}) as
\begin{equation} \label{ChemicalParameters}
w_0  =  1-n+w_2, \ \ \  
w_{\uparrow}  =  w_{\downarrow} = \frac{n}{2} - w_2.  
\end{equation}
The value of $w_2$ can again be extracted from the ground state energy via the relation
$w_2 = (\partial E_0 / \partial u)/4L$, where the {\em Bethe Ansatz} solution
for $E_0$ can now be expressed via a $1/u$ expansion \cite{Carmelo}:  
\begin{eqnarray} \label{LargeExpansion}
\frac{E_0}{L}&=&-\frac{2}{\pi}\sin(\pi n)-\sum_{l=1}^{\infty}\kappa_l(n)\left(\frac{1}{4u}\right)^l.
\end{eqnarray}
The values of $\kappa_l(n)$ are tabulated to fifth order in Ref. \cite{Carmelo}. 
The ground state energy in (\ref{LargeExpansion}) also determines the chemical potential as function of
filling:
$\mu(n) = \partial  E_0/\partial n$. By inverting $\mu(n)$ and inserting the resulting values
for the $w$-parameters from (\ref{Parameters}) into (\ref{OneTangle}) we can plot ${\cal E}$ vs.
$\mu$
for any value of $u\!>\!1$. Some representative plots are shown in Fig.
(\ref{ChemPot}), 
together with the single-site entanglement for free electrons ($u=0$).

In order to analytically explore the quantum critical regions $\mu \rightarrow \mu_{c1 +}$ and $\mu
\rightarrow \mu_{c2 -}$
we first consider the $u \rightarrow \infty$ limit where
$w_2=0$. In this limit (\ref{LargeExpansion}) implies that
$n(\mu) = (1/\pi) \arccos (-\mu / 2)$.
Combining this expression with Eqs. (\ref{OneTangle}) and (\ref{ChemicalParameters}) we obtain 
\begin{eqnarray} \label{RepulsiveDerivative}
\frac{\partial {\cal E}}{\partial \mu} =
(-1)^i\frac{\chi_{Ci}}{2\ln (2)}(\ln |\mu-\mu_{ci}|+\mbox{const}.), \ i\!=\!1,2
\end{eqnarray}
for $\mu \rightarrow \mu_{c1 +}$ and $\mu \rightarrow \mu_{c2 -}$, respectively.
Note that the derivative of the single-site entanglement is again given by a susceptibility, corrected by a logarithmic
factor 
that reflects the change in the number of available local states 
as the system undergoes the transition: When $\mu \rightarrow \mu_{c1 +} \
[\mu \rightarrow \mu_{c2 -}]$ the singly occupied [empty] local states get suppressed.
(Cf. the argument after Eq. \!(\ref{OneTangleGradient1}).) The exact analogy with the magnetic scaling
in (\ref{OneTangleGradient1}) can be understood 
by carrying out a
particle-hole transformation for spin-up electrons: $\cdop{j \uparrow} \leftrightarrow (-1)^j \cop{j
\uparrow}$ (leaving the spin-down electrons untouched).
This transformation maps the zero-field attractive system with a variable chemical potential 
onto a half-filled repulsive system with a variable magnetic field.
It follows    
that the single-site entanglement at $\mu_{c1} \ (\mu_{c2})$ has the same behavior as at $h_{c2} \ (h_{c1})$.

Turning to the case of large but finite $u$, we focus on the Mott transition $\mu \rightarrow \mu_{c2 -}$.
A straightforward analysis, again using the {\em Bethe Ansatz} result in (\ref{LargeExpansion}), yields for the
leading behavior
of the single-site entanglement:
\begin{equation}  \label{FiniteU}
\frac{\partial{\cal E}}{\partial\mu} =  - C(u) \chi_{C2},  
\end{equation}
with $C(u)$ a positive $u$-dependent constant.
Note that there is no logarithmic correction in (\ref{FiniteU}): When $u$ is finite the metallic
($\mu\!<\!\mu_{c2}$)
and insulating ($\mu\!>\!\mu_{c2}$) ground states for $0\!\le\!n\!\le\!1$ are both superpositions 
of all four types of local states
$\ket{0}_j, \, \ket{\,\uparrow}_j, \,\ket{\,\downarrow}_j$, and $\ket{\,\uparrow \downarrow}_j$. It
follows that none of
the $w$-parameters in (\ref{OneTangle}) tend to zero, and the logarithmic terms
add up to a constant \cite{footnote}. 

The results in (\ref{RepulsiveDerivative}) and (\ref{FiniteU})
derived for $0\!\le\!n\!\le\!1$ can be extended to the region $1\!\le\! n\!\le\!2$ via the 
particle-hole transformation
$\cdop{j \alpha} \leftrightarrow (-1)^j \cop{j \alpha}, \alpha = \uparrow, \downarrow$. One finds that the
change of ${\cal E}$ at the quantum critical point
$\mu_{c3}=4u-\mu_{c2}$ [$\mu_{c4}=4u-\mu_{c1}$] where the system goes to more than half-filling, $n>1$
[complete filling,
$n=2$] exhibits the same scaling as at the transition to half-filling [empty lattice] studied above.

To summarize, we have found that the derivatives $\partial {\cal E}/\partial h$ and $\partial
{\cal E}/\partial \mu$
of the single-site entanglement ${\cal E}$ are 
faithful markers for QPTs in the Hubbard model
driven by a change in magnetic field $h$ or chemical potential $\mu$. Via an analysis based on the {\em
Bethe Ansatz}
solution of the model, we have derived exact expressions for $\partial {\cal E}/\partial h$ and $\partial
{\cal E}/\partial
\mu$ at criticality, revealing that these quantities scale with the corresponding diverging spin and
charge susceptibilities,
respectively. Logarithmic corrections signal a change in the number of available local states at the QPT.
That a critical entanglement entropy is directly connected to a susceptibility is an
intriguing property.
A zero-temperature susceptibility is an observable determined only by the dependence of the ground state 
energy on the critical
parameter (magnetic field or chemical potential). The entanglement entropy, on the other hand, carries information
about the very structure of the 
ground state. From our analysis it is transparent how the linkage 
{\em formally} comes about for the Hubbard model: The
reduced density
matrix can be parameterized by expectation values derivable from the ground state energy via the
Hellman-Feynman theorem.
To find a physical interpretation of the connection, and to explore whether it can be extended to other
critical quantum
many-particle systems, is an interesting and challenging problem. \\ 
       
We thank R. Rehammar and O. Sylju\aa sen for valuable discussions. 
We are also grateful to K. Capelle and V. V. Fran\c ca for drawing our attention to a numerical error in 
the original Fig. 2 of this paper (published in Phys. Rev. Lett. {\bf 95}, 196406 (2005)) \cite{erratum}. 
H.J. acknowledges support from the
Swedish
Research Council.


\begin{thebibliography}{99}
\bibitem{Note}
Note that a pure state measure of entanglement is still applicable at zero temperature
when there is no dynamic interaction with the environment.
\bibitem{footnote}
In Ref. \cite{Gu}, 
the change of
the local entanglement of the 1D Hubbard model was found to
exhibit a finite discontinuity at the Mott transition at $n=1$, and it was
argued that this
is a characteristic of the transition.
It is important to note, however, that the discontinuity at $n=1$ is a mere consequence of the fact that
the limits $n \rightarrow 1^{-}$ and $n \rightarrow 1^{+}$ correspond to distinct regions for the chemical potential,
$\mu \rightarrow \mu_{c2 -}$ and $\mu \rightarrow \mu_{c3 +}$, respectively.
In fact, what matters is not the change of the entanglement with $n$, but with $\mu$
(which is the experimentally controllable parameter).
As shown by our exact results, $\partial{\cal E}/\partial \mu$
exhibits {\em divergences} at $\mu_{c2}$ and $\mu_{c3}$, suggesting that the proposed
classification of second order QPTs by Wu {\em et al.} \cite{Wu} may be extended to
local entanglement measures for fermionic systems.
\end{thebibliography}

\end{document}